\documentclass[final,5p,times,twocolumn]{elsarticle}

\usepackage{graphicx}

\usepackage{amssymb}

\usepackage{lineno}

\usepackage{subfigure}
\usepackage{booktabs}
\usepackage{multirow}
\usepackage[permil]{overpic}
\usepackage[load-configurations=abbreviations]{siunitx}
\usepackage{upgreek}
\usepackage{xcolor}

\journal{Nuclear Instruments and Methods A}

\begin{document}

\begin{frontmatter}

\title{Upgrade of the ALICE ITS detector}

\author{F.~Reidt \\ {\small on behalf of the ALICE collaboration}}
\address{CERN, 1211 Geneva 23, Switzerland}
\ead{felix.reidt@cern.ch}

\begin{abstract}
The ALICE experiment at CERN has undergone a major upgrade during the Long Shutdown 2 of the LHC during 2019-2021.  A key element is the installation of the new Inner Tracking System 2 (ITS2). The new ITS2 based on CMOS Monolithic Active Pixel Sensors will significantly improve the impact parameter resolution and the tracking efficiency, especially for particles with low transverse momenta, as well as the readout-rate capability. In the following, the upgrade and its commissioning status as well installation status will be outlined.
\end{abstract}

\begin{keyword}
Silicon Pixel Detector \sep MAPS \sep ALPIDE \sep Vertex Detector
\end{keyword}

\end{frontmatter}

\section{Introduction}
\label{}
The ALICE experiment~\cite{Aamodt2008} has been upgrading its experimental apparatus during the currently ongoing Long Shutdown~2 (LS2) of the LHC during 2019 to 2021.
This upgrade~\cite{Alice2014a} rests on two pillars: the recording Pb--Pb collision data at \SI{50}{kHz} interaction rate to increase the minimum-bias data sample by about a factor of one hundred, the increased vertexing capabilities while maintaining previous particle identification performance. This is motivated by high-precision measurements of rare probes at low transverse momentum ($p_\textrm{T}$).

\section{Inner Tracking System 2 (ITS2)}
The upgraded Inner Tracking System 2 (ITS2) is a key element of upgrade of the experimental apparatus of the ALICE experiment.  The main goals of the ITS2 upgrade are to achieve an improved reconstruction of the primary vertex as well as decay vertices originating from heavy-flavour hadrons, and an improved performance for the detection of low-$p_\textrm{T}$ particles. The design objectives of ITS2 are the following ones:
\begin{enumerate}
\item  The impact parameter resolution will improve by a factor of 3 and 5 in the $r\varphi$ and $z$ coordinate, respectively, at a $p_\textrm{T}$ of $\SI{500}{\MeV/\mathit{c}}$ by three main changes:
\begin{itemize}
\item moving the detector closer to the interaction point from previously \SI{39}{mm} to \SI{23}mm for the innermost layer;
\item reducing the material budget of the inner layers to \SI{0.35}{\percent~X_{0}} compared to \SI{1.14}{\percent~X_{0}} of its predecessor;
\item reducing the pixel size from
  $\SI{50}{\micro\meter}\times\SI{425}{\micro\meter}$ to \mbox{$\SI{29.24}{\micro\meter}\times\SI{26.88}{\micro\meter}$}.
\end{itemize}
\item The tracking efficiency and the $p_\textrm{T}$ resolution at low $p_\textrm{T}$ will improve through the increased granularity of the 7 layer, fully pixelated detector.
\item The readout rate will be increased from \SI{1}{kHz} up to \SI{100}{kHz} in \mbox{Pb--Pb} and \SI{200}{kHz} in pp collisions.
\end{enumerate}

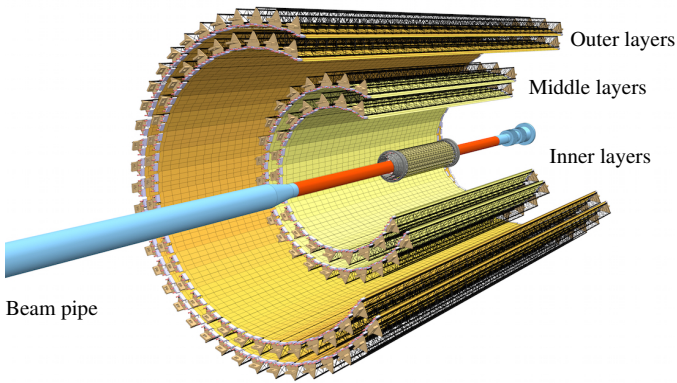
\begin{figure}[b]
  \begin{overpic}[width=0.5\textwidth]{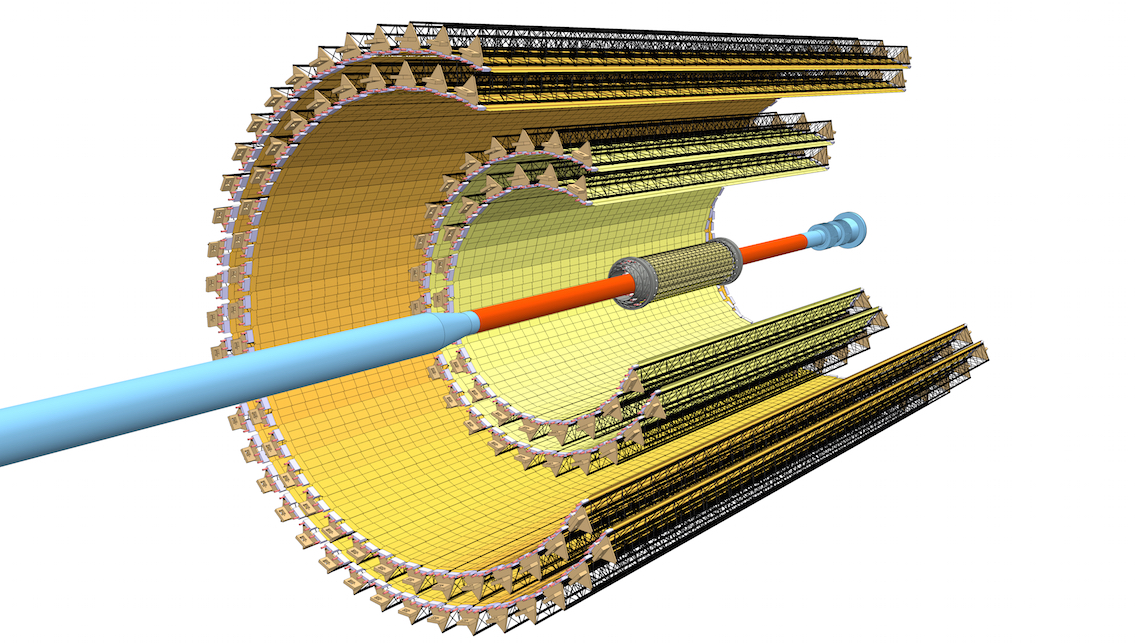}
    \put(   0,100){\footnotesize Beam pipe}
    \put( 810,490){\footnotesize Outer layers}
    \put( 750,420){\footnotesize Middle layers}
    \put( 780,320){\footnotesize Inner layers}
  \end{overpic}
  \caption{ITS2 layout, taken from~\cite{Alice2014b}.\label{fig:ITSlayout}}
\end{figure}

The layout of the detector is depicted in Fig.~\ref{fig:ITSlayout}. The radii of the concentric layers range from \SIrange{23}{400}{\milli\meter}. A total of 192 staves form an active silicon area of \SI{10}{\meter\squared} amounting to \SI{12.5e9}{pixels}.  The innermost three layers form the so-called Inner Barrel (IB) containing 48 staves of \SI{27}{\centi\meter} length. The Outer Barrel (OB) consists of the Middle Layers (MLs) and Outer Layers (OLs) which are assembled from 54 and 90 staves of \SI{84}{\centi\meter} and \SI{150}{\centi\meter}, respectively.

\begin{figure}
  \includegraphics[width=0.5\textwidth]{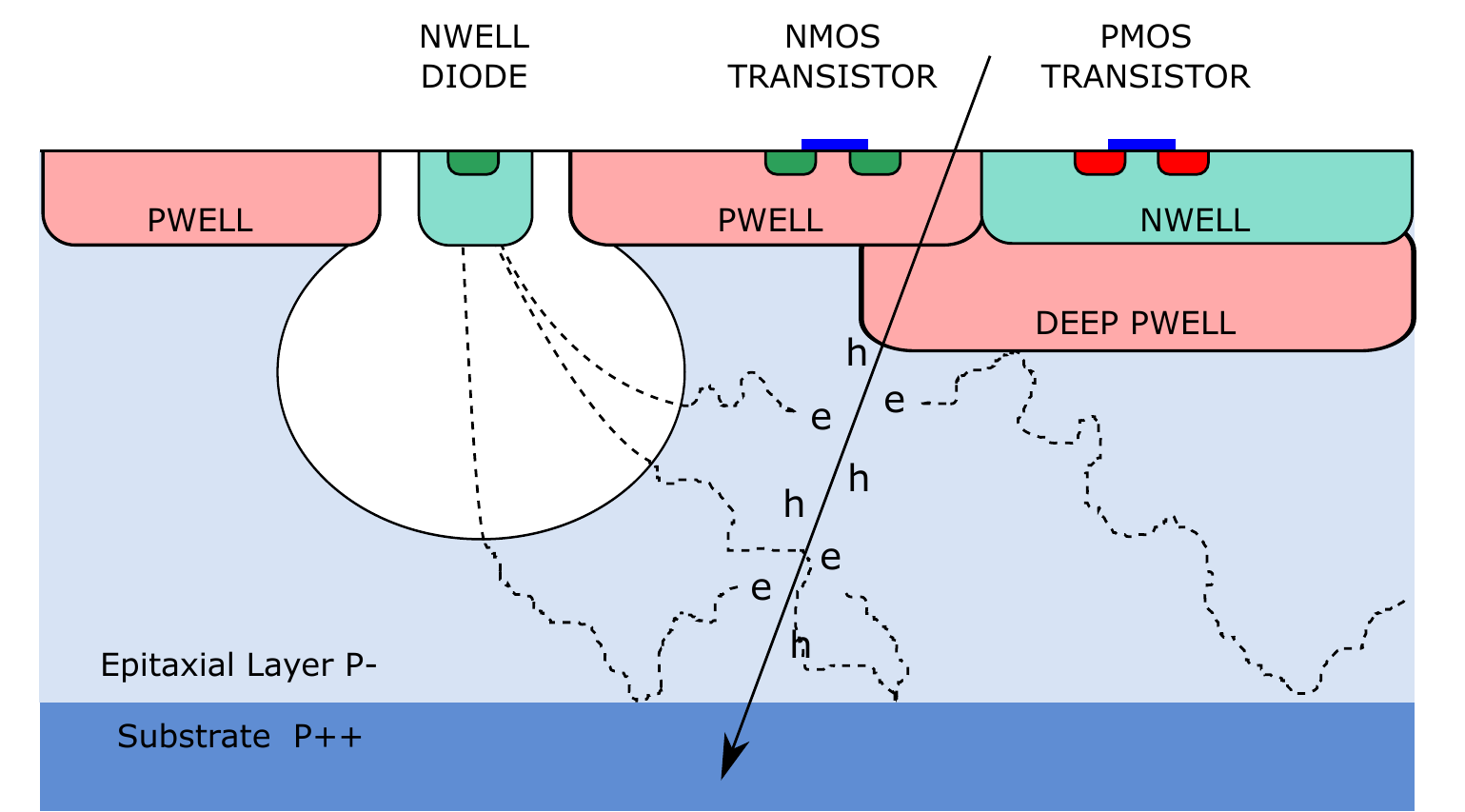}
  \caption{Schematic cross section of a pixel of a monolithic active silicon pixel sensor using the TowerJazz CMOS Imaging process, taken from~\cite{Alice2014b}.\label{fig:MAPS}}
\end{figure}

The principal component of the ITS2 is the ALPIDE chip~\cite{AglieriRinella:2017lym}, a Monolithic Active Pixel Sensor (MAPS) fabricated in the TowerJazz \SI{180}{\nano\meter} CMOS Imaging Process. A deep p-well allows to employ complex logic inside the pixel matrix without having the n-wells PMOS transistors compete with the collection electrodes (cf.~Fig.~\ref{fig:MAPS}). Therefore, ALPIDE features in-pixel amplification, shaping and discrimination as well multiple-event buffers. A priority-encoder circuit is employed to read only the addresses of hit-pixels and achieve in-matrix data-sparsification. The key characteristics of ALPIDE are summarised in Tab.~\ref{tab:ITSrequirements}.

\begin{table}[t]
  \caption{ALPIDE  pixel-chip characteristics.}\label{tab:ITSrequirements}
  \centering
  \begin{tabular}{lc}
    \toprule
    \textbf{Parameter}   & \textbf{Value} \\ \midrule
    Chip dimensions       & $\SI{15}{\mm}\times\SI{30}{\mm}$ ($r\varphi\times z$) \\
    Spatial resolution      & \SI{5}{\um}      \\
    Detection efficiency  & $>$\,\SI{99}{\%} \\
    Fake-hit rate              & $\ll$\SI{1e-6}{\per event \per pixel} \\
    Power density           & $<$ \SI{40}{\mW/\cm\squared} \\
    TID radiation hardness  & $>$\SI{270}{krad} \\
    NIEL radiation hardness & $>$\SI{1.7e12}{\SI{1}{MeV~n_{eq}/\cm\squared}} \\ \bottomrule
  \end{tabular}
\end{table}

The ITS2 provides pseudo-rapidity coverage of $|\eta|<\num{1.22}$ for $\SI{90}{\%}$ of the most luminous beam interaction region. The radial positions of the layers were optimised in order to achieve the best combined performance in terms of pointing resolution, $p_\textrm{T}$ resolution and tracking efficiency in Pb--Pb collisions at hit densities of about \SI{10}{\cm^{-2}\per event} on average for minimum-bias events in the innermost layer. The ITS2 will be operated at room temperature (\SIrange{20}{25}{\degreeCelsius}) using water cooling. The expected radiation load at the innermost layer amounts to \SI{270}{krad} of Total Ionising Dose (TID) and \SI{1.7e12}{\SI{1}{MeV~n_{eq}/\cm\squared}} of Non-Ionising Energy Loss (NIEL). In order to meet the material budget requirements, the silicon sensors for the innermost layers are be thinned down to \SI{50}{\um}.

\begin{figure}
  \includegraphics[width=0.5\textwidth]{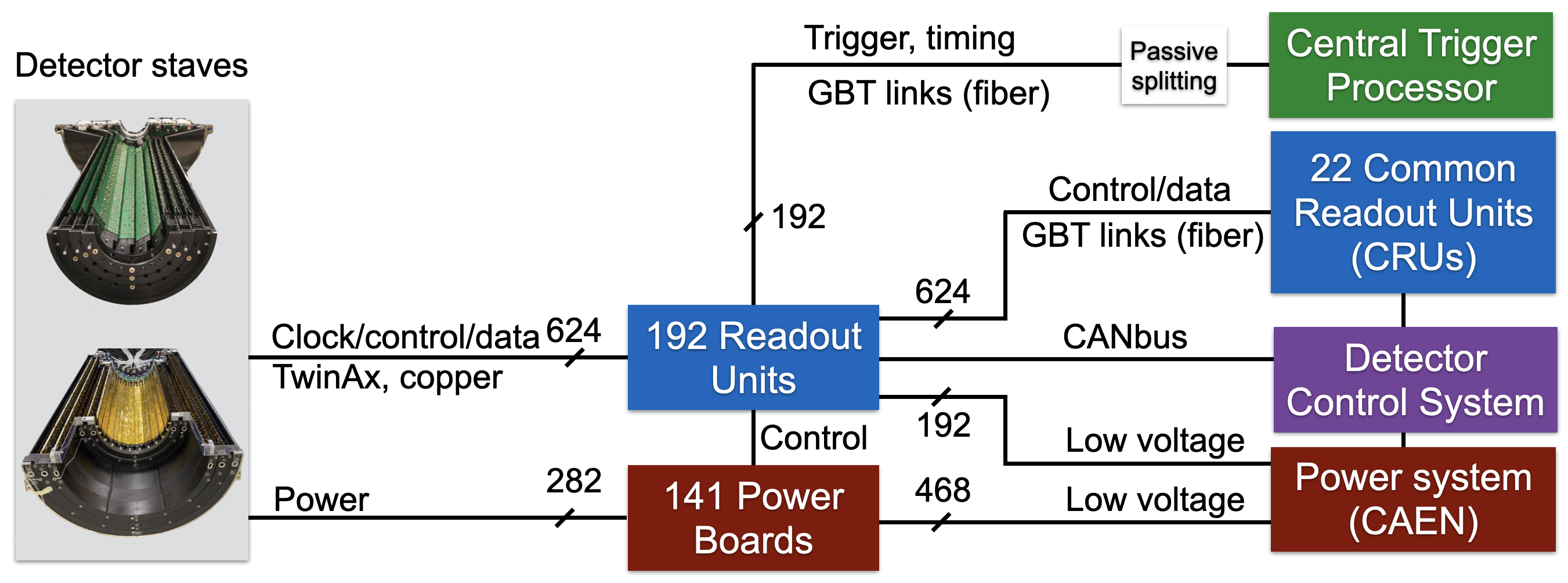}
  \caption{ITS2 system overview.\label{fig:ITSsystem}}
\end{figure}

The ITS2 system is schematically shown in Fig.~\ref{fig:ITSsystem}. The detector staves are connected to the Readout Units (RUs) via roughly \SIrange{6}{8}{\meter} long TwinAx copper cables providing clock, control and data transport. In the IB, every ALPIDE chip transmits data via a dedicated \SI{1.2}{Gbit\per s} high-speed link. In the OB, seven ALPIDE chips share a \SI{400}{Mbit\per s} link which is possible due to the lower occupancy in the middle and outer layers. The RUs control the Power Boards (PBs) which supply the detector via \SIrange{6}{8}{\meter} long power cables. RUs and PBs are located inside a magnetic field of \SI{0.5}{T} and are exposed to less than \SI{10}{\kilo rad} of TID. The RUs are connected to the the upstream Common Readout Units (CRUs) by more than 600 GigaBitTransceiver (GBT) links. The CRUs are PCIe-based interface cards located in the First-Level-Processing (FLP) servers. Trigger and timing information is supplied to the RUs from the ALICE Central Trigger Processor (CTP) via a single-mode fiber network employing passive optical splitting into almost 200 GBT links. The main control path for the Detector Control System (DCS) is via the CRU and the GBT links. CANbus serves as slow backup communication interface which can be used to do status monitoring or control the power towards the detector.

The detector construction and assembly started in December 2016 with the ALPIDE mass production. A total of more than \num{72000} ALPIDE chips were assembled into roughly \num{2600} Hybrid Integrated Circuits (HICs) consisting of Flexible Printed Circuits and ALPIDE chips which in turn were mounted on detector staves.
An IB HIC combines nine ALPIDE chips which are mounted together on a stave. In the OB, a stave consists of 8 or 14 HICs for installation in ML and OL, respectively. An OB HIC combines two rows of seven chips.

The full assembly of the detector half-barrels continued until the end of 2019. The ITS production is summarized e.g. in~\cite{Contin:2020cbu}.

\section{On-Surface Commissioning}
With the availability of first IB half layers in May 2019, the commissioning of the detector in a cleanroom on surface started and further components were integrated step-by-step.

In order to provide testing with a system as final as possible, the services to be used in the experiment, including cooling plant, power system and computing system for detector control and data acquisition, were first installed on the surface and then later moved together with the detector for final integration and commissioning in ALICE.
Throughout 2020, the full detector was under commissioning on-surface with the half-barrels located next to each other to facilitate step-wise integration.

\begin{figure*}
  \centering
  \includegraphics[width=0.95\textwidth]{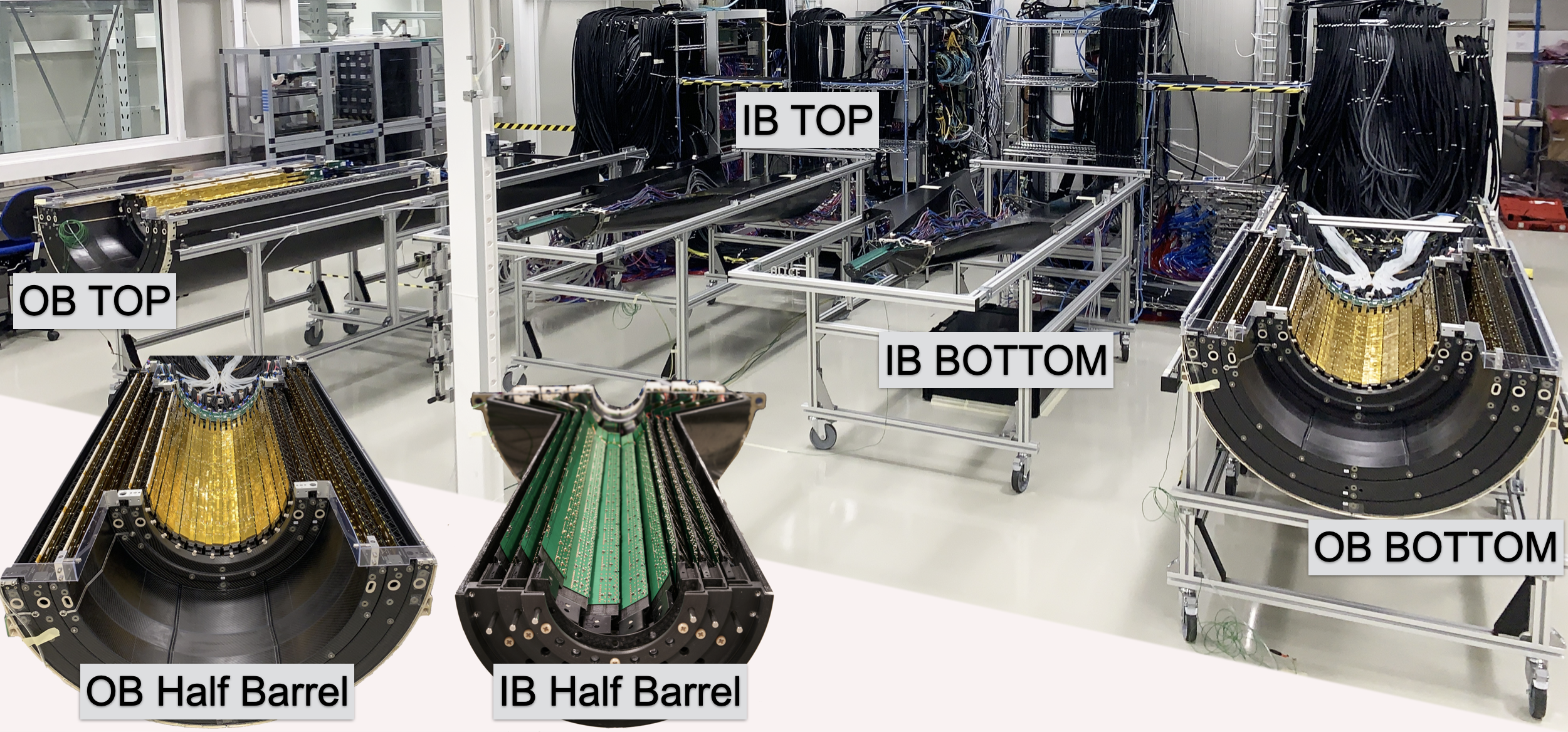}
  \caption{ITS2 in the clean room during on-surface commissioning, in the backfrom left to right OB top half-barrel, IB top half-barrel, IB bottom half-barrel, OB bottom half-barrel; zoomed view of an OB half-barrel (left) and an IB half-barrel (right).\label{fig:ITSonSurface}}
\end{figure*}

During the on-surface commissioning, the performance of the detector was thoroughly characterised with regard to its detection, noise and readout performance. Furthermore, the charge threshold of the detector was calibrated to achieve an average value of \SI{100}{\elementarycharge^{-}} which was used to obtain the results shown in this section. This working point chosen for the operation of the ITS based on previous measurements~\cite{Martinengo:2017fuc}.

\subsubsection*{IB Readout Performance}
The performance of the ALPIDE high-speed link was verified for various combinations of occupancies and readout rates. The occupancy was generated by pulsing clusters of eight pixels with a static pattern being repeated in every event. The amount of clusters was varied from 32 to 512 clusters. In comparison, the average occupancy from minimum bias \mbox{Pb--Pb} collisions is expected to be below 32 clusters, and 64 clusters correspond to a most-central \mbox{Pb--Pb} collision. An upper limit for the bit-error rate was set to \num{1e-16} for an occupancy of 64 clusters up to a rate of \SI{247}{kHz}. A comparable perfomance was confirmed at a rate of \SI{44.9}{kHz} increasing the occupancy of 512 clusters. This translates into a large margin with respect to occupancy and readout rate and would allow for several tens of hours of operation of the full system without any bit error at nominal conditions.

\begin{figure}
  \centering
  \includegraphics[width=0.45\textwidth]{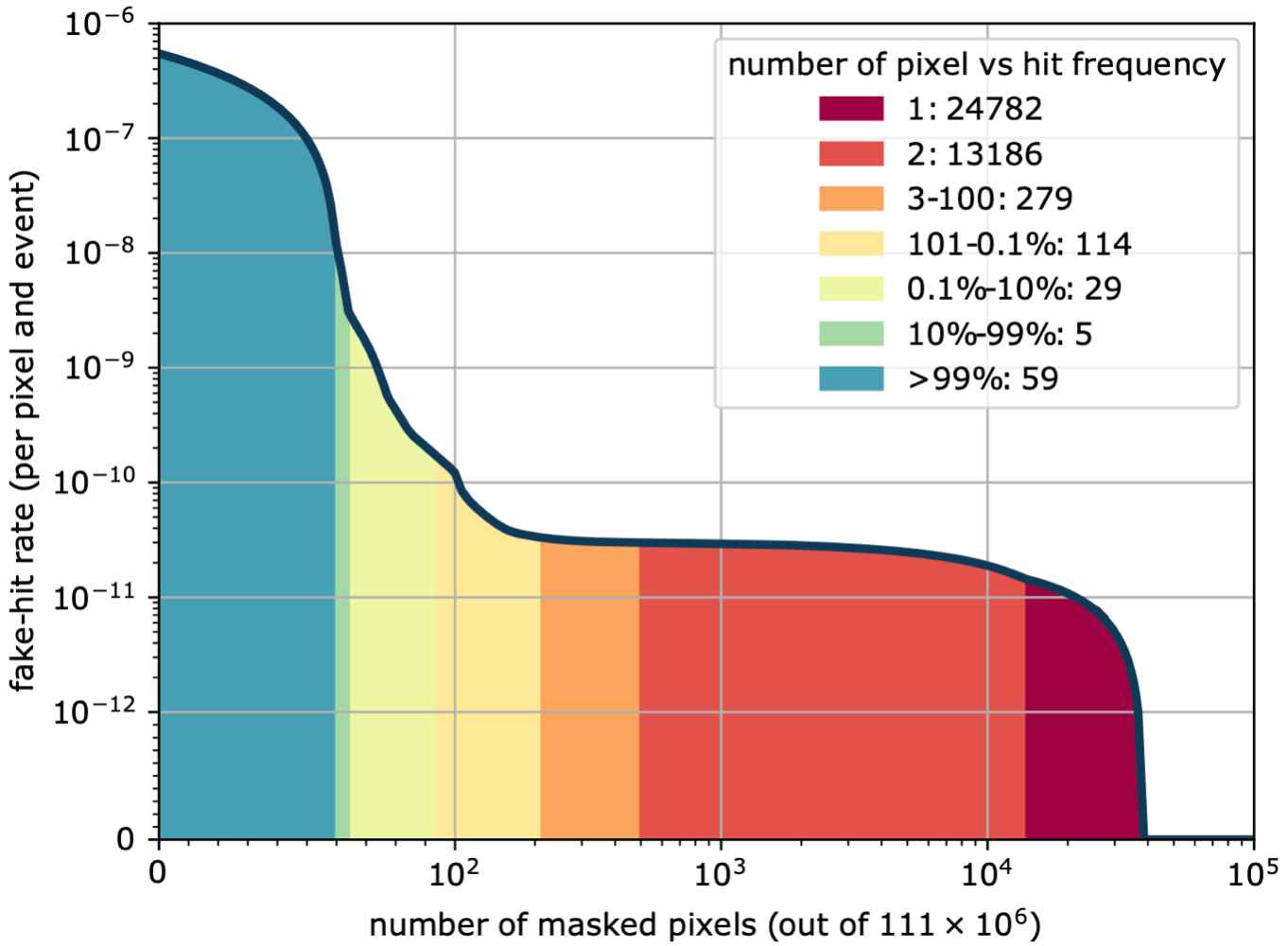}
  \caption{The fake-hit rate of a IB Half-Barrel as function of the number of masked pixels, colors indicate how often a pixel fired in \SI{15e6}{events} acquired at a trigger rate of \SI{50}{kHz} using a charge threshold of \SI{100}{\elementarycharge^{-}} .\label{fig:IBfhr}}
\end{figure}

\subsubsection*{IB Fake-Hit Rate}
\begin{figure}
  \centering
  \includegraphics[width=0.5\textwidth]{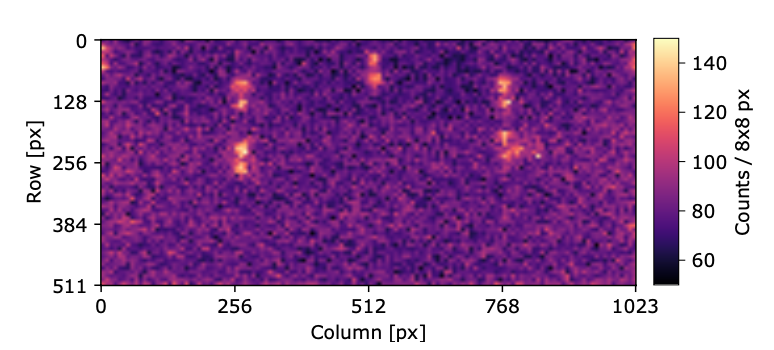}
  \includegraphics[width=0.35\textwidth]{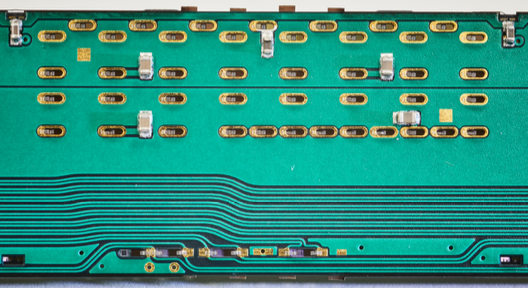}
  \caption{Superimposed hit-maps of 216 ALPIDEs amounting to 204 days of exposure time (top) and photograph of an IB HIC (bottom), taken from \cite{Mager2021}.\label{fig:IBpb}}
\end{figure}

In the IB, the fake-hit rate was found to be dominated by roughly hundred pixels per half-barrel as shown in Fig.~\ref{fig:IBfhr}. Masking these pixels leads to a fake-hit rate of \SI{1e-10}{\per pixel \per event}. This is significantly better than the required \SI{1e-6}{\per pixel \per event }. The majority of pixels which remains after this masking shows one or two hits, which are compatible with the hypothesis of being hits from cosmic muons. After subtraction of hits which can be associated to a track and overlaying the hits of all chips on top of each other, the hit-map in Fig.~\ref{fig:IBpb} (top) was obtained. By comparing the hit-map to a photograph of the IB HIC in Fig.~\ref{fig:IBpb} (bottom), one can clearly associate noise hit clusters to the position of the decoupling capacitors. A dedicated study~\cite{Mager2021} showed these hits to originate from electrons emitted in the decay chain of Pb-210 which was part of the solder used to assemble the HICs.

\subsubsection*{OB Fake-Hit Rate}
\begin{figure}
  \includegraphics[width=0.45\textwidth]{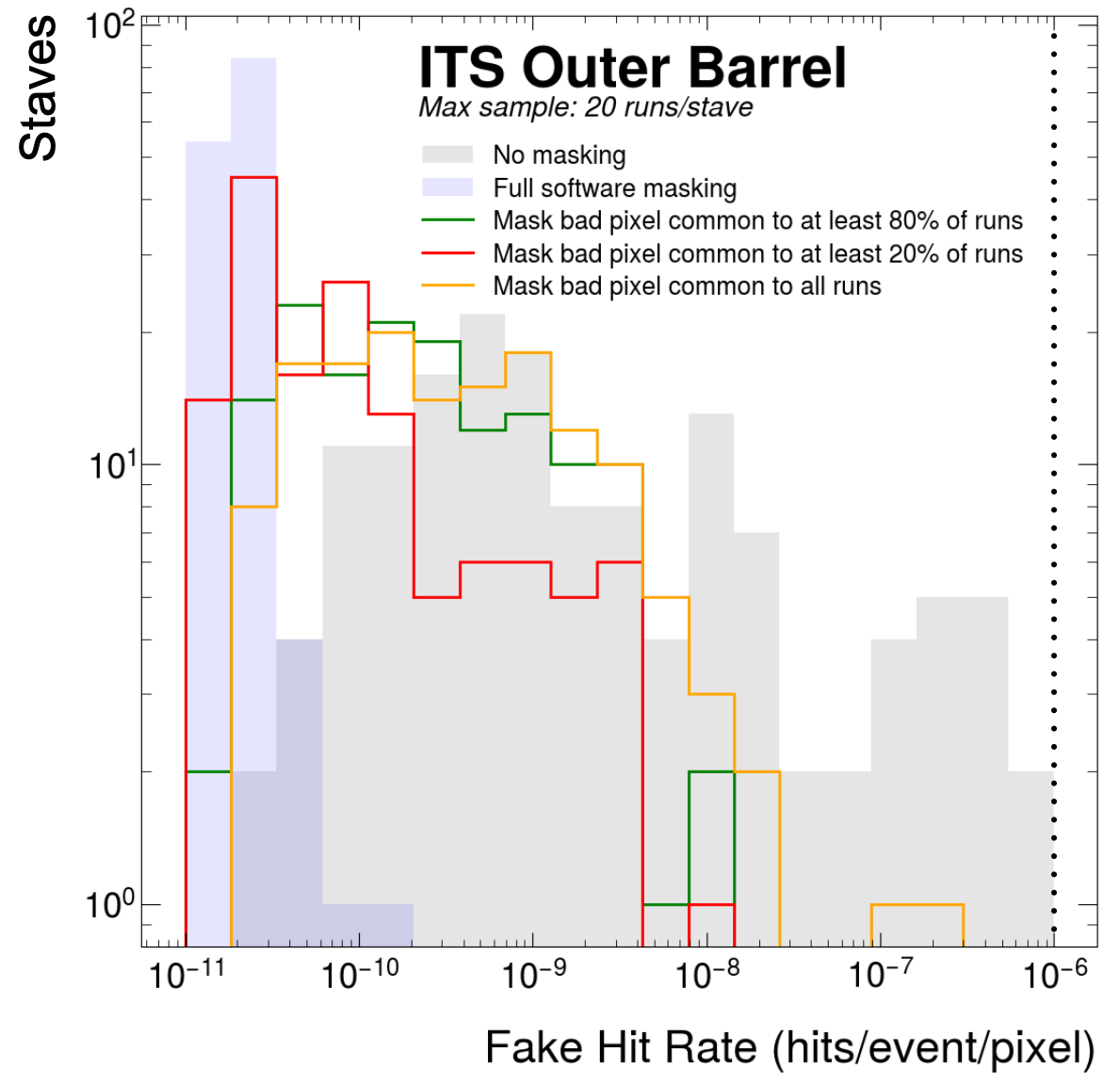}
  \caption{OB fake-hit rate for different masking conditions.\label{fig:OBfhr}}
\end{figure}
The noise performance has also been confirmed for the Outer Barrel of the ITS2. After masking of broken double columns, all chips show a fake-hit rate below \SI{1e-6}{\per pixel \per event} (cf. Fig~\ref{fig:OBfhr}, solid grey). This figure is based on a minimum of twenty noise acquisition runs of approximately one million events.

A realistic estimate of the performance of the online masking is given by masking pixels which are noisy in all runs. This leads to a significant reduction of the fake-hit rate and the vast majority of staves exhibits a value below \SI{1e-8}{\per pixel \per event}  (cf., Fig.~\ref{fig:OBfhr}, yellow). Compared to the required \SI{1e-6}{\per pixel \per event}, this represents a significant decrease of data volume from noise which dominates in the outer layers and in particular for pp collisions. The reason for the variation of which pixels are noisy over time is to be understood. A small fraction of the noisy pixels could be simply showing thermal noise, however, it is unlikely to explain the full extent. Another contribution could be variations in the supply voltage which are particularly pronounced in the OB as here up to 98 chips on a half-stave share the same return path.

Offline masking even allows one to obtain a fake-hit rate of \SI{1e-10}{\per pixel \per event} (cf.light blue), however, this can only be exploited during the processing of the data.

\subsubsection*{OB Tracking Efficiency}
The excellent noise performance furthermore allows to efficiently acquire cosmic muon tracks using continuous integration without employing a dedicated trigger. For the OB, these tracks were used to study the detection efficiency. The clusters of three layers were fitted with a straight line and extrapolated to the fourth layer. Preliminary results are around \SI{99}{\percent} for the detection efficiency of all four layers of the OB. This study has been carried out without software alignment, relying on the mechanical alignment of the detector mechanics. 
%

\section{Installation in the ALICE Experiment}
\begin{figure}
  \centering
  \includegraphics[width=0.45\textwidth]{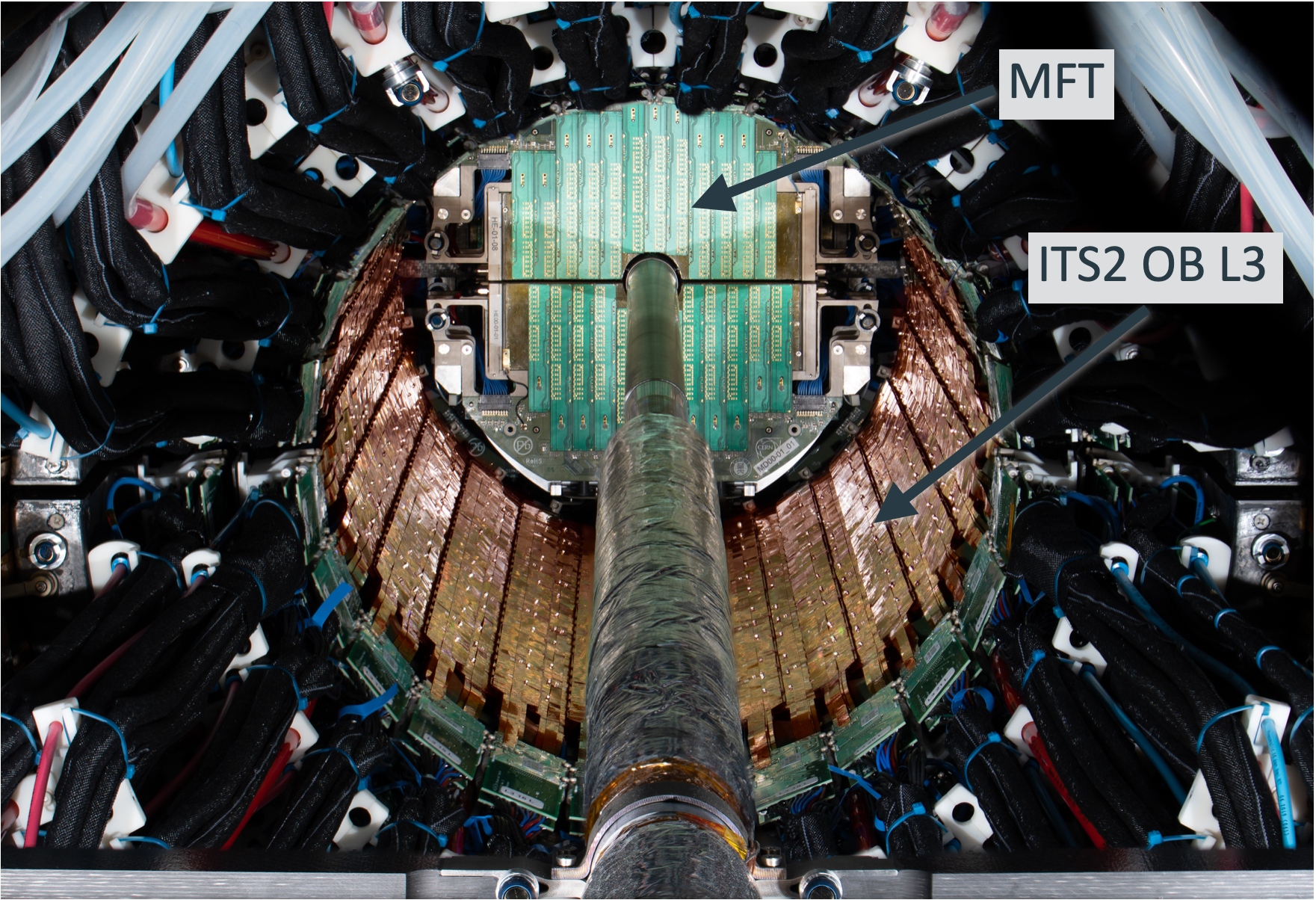}
  \includegraphics[width=0.45\textwidth]{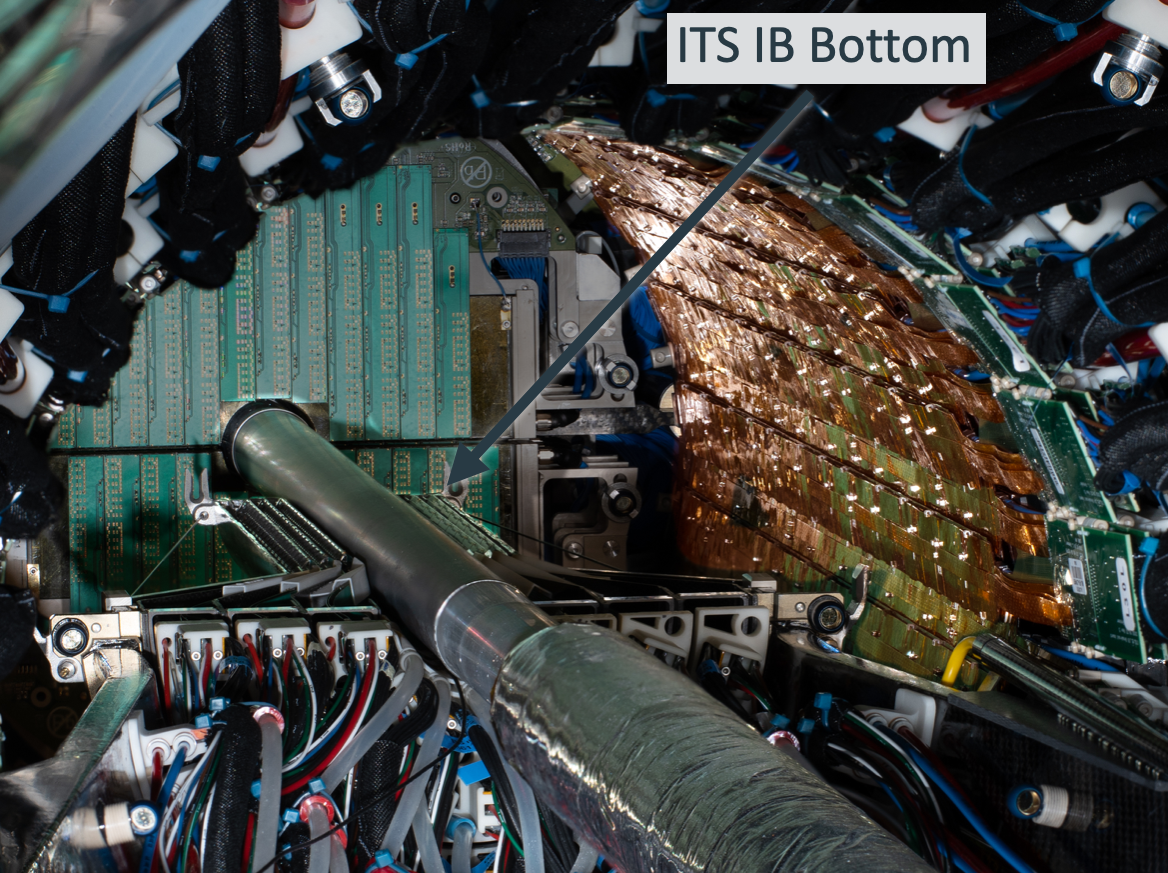}
  \caption{OB surrounding the beam pipe with the Muon Forward Tracker (MFT) in the background (top), \label{fig:OBinstallation} ITS2 IB bottom half-barrel next to the beam pipe, OB and MFT in the background.\label{fig:IBinstallation}}
\end{figure}

At the beginning of 2021, the services were transferred from the on-surface commissioning hall to the experimental cavern. In March 2021, the Outer Barrel was installed around the interaction point. Figure~\ref{fig:OBinstallation} (top) shows Layer 3, and in particular its power distribution bus shining golden, around the beam pipe. After the installation, the detector was thoroughly verified before the Inner Barrel was inserted into its final position in close vicinity of the beam pipe. In Fig.~\ref{fig:IBinstallation} (bottom), the bottom half of the Inner Barrel is in its final position at around \SI{1}{mm} distance from the beam pipe.

ITS2 is installed on rails in the so-called cage hosting beam pipe, ITS2 and MFT~\cite{Alice2015c} inside the ALICE TPC~\cite{ALICETPC:2020ann}.
Due to the small clearances and the fragile nature of the detectors, the ITS2 installation was highly challenging. Therefore, several insertion tests have been carried out on surface to optimise the procedures and to spot potential interferences. For the final installation inside the experimental apparatus, up to six cameras where used to monitor key contact points. Furthermore, surveys were carried out to determine the exact position of the detector elements and the beam pipe and visualize them with the help of three-dimensional scans which had be carried out before. During the insertion process moreover CAD models with increasing insertion progress were compared to the camera images to verify the positioning of the detector elements. The full detector was installed without damaging any component.

\section{First Results from ITS2 in the ALICE Experiment}
After connection and verification of the detector and its services, the focus was on the central system integration and on gaining experience with the final framework for the operation of the detector.

\begin{figure}
  \centering
  \includegraphics[width=0.45\textwidth]{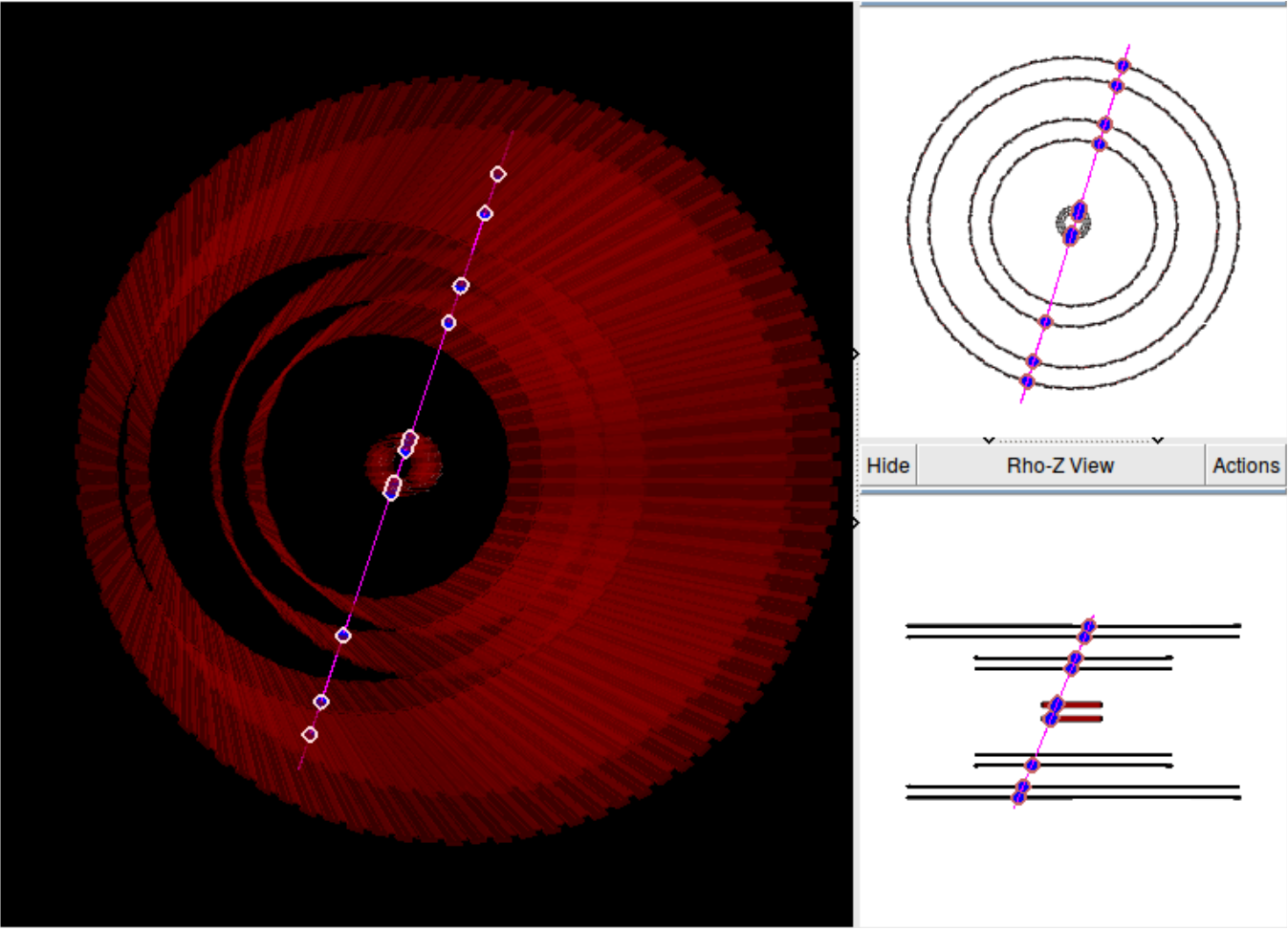}
  \caption{Cosmic muon traversing all layers of ITS2 twice, no magnetic field.\label{fig:ITScosmic}}
\end{figure}

\begin{figure}
  \centering
  \includegraphics[width=0.45\textwidth]{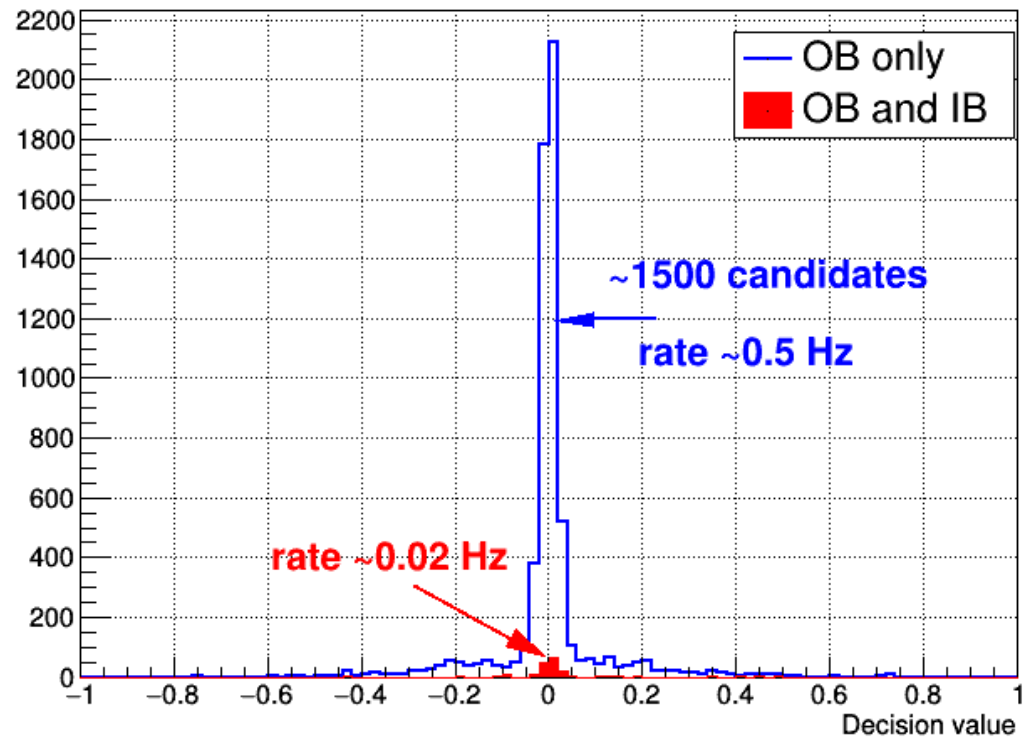}
  \caption{Track straightness based on hits in layers 4, 5 and 6 in the OB for tracks passing only the OB (blue) as well as both IB and OB (red).\label{fig:DecisionValue}}
\end{figure}

Furthermore, first cosmic muon tracks like the one shown in Fig.~\ref{fig:ITScosmic} traversing the full detector could be acquired using continuous integration without a dedicated trigger signal. These tracks were found by matching three hit points in the layers 4 to 6 of the OB and requiring another hit point in the IB. Such tracks were observed at a rate of about \SI{0.02}{Hz}, while tracks traversing only the OB were more frequent, with a rate of around \SI{0.5}{Hz}. Figure~\ref{fig:DecisionValue} shows the so-called decision value indicating the straightness of the seeding tracklet in layers 4 to 6 for tracks passing through OB only (blue line) as well as tracks traversing both IB and OB (red).


The study of these first tracks to determine detection efficiency and alignment is currently ongoing. The residuals prior to alignment are found to be of the order of \SI{1}{\milli\meter} underlining the excellent mechanical alignment and confirming the correctness of the mapping of the detector channels in software

A first look at the noise performance using a coarse tuning of the charge threshold was possible via the newly developed ITS2 Quality Control (QC). At this stage of the commissioning inside the experimental apparatus, the majority of IB staves showed already fake-hit rates around \SI{1e-8}{\per pixel \per event}. For the OB, most of the staves showed a fake-hit rate better than \SI{1e-6}{\per pixel \per event}.

During the remaining commissioning time, the tools for calibration and detector control will be finalised and the settings of the detector will be optimised to match the performance observed during the on-surface commissioning.

\section{Summary}

The upgrade Inner Tracking System, ITS2, has been installed in the experimental apparatus. Results from the on-surface commissioning as well as of the commissioning inside the experimental apparatus confirm the performance of single ALPIDE chips with regard to fake-hit rate and detection efficiency. The second half of 2021 is devoted to global commissioning together with the ALICE sub-detectors and first data with beam provided by the LHC pilot beams in October 2021 in preparation for the LHC Run 3 in 2022.


\bibliographystyle{elsarticle-num}
\bibliography{literature}

\end{document}